\documentclass[twocolumn,showpacs,preprintnumbers,amsmath,amssymb,nofootinbib]{revtex4-1}

\def\be{\begin{equation}}
\def\ee{\end{equation}}

\def\be{\begin{equation}}
\def\ee{\end{equation}}
\def\beq{\begin{equation}}
\def\eeq{\end{equation}}

\def\be{\begin{equation}}
\def\ee{\end{equation}}

\usepackage[dvipdfmx]{graphicx}
\usepackage{here}
\usepackage{comment,braket}
\usepackage{epsf}
\usepackage{amsmath}
\usepackage{graphics}
\usepackage{amsfonts}
\usepackage{amssymb}
\usepackage{latexsym}
\usepackage{color}
\usepackage{ulem}
\input{colordvi.tex}
\usepackage{feynmp}

\usepackage{bm}
\usepackage[svgnames]{xcolor}
\definecolor{phthaloblue}{rgb}{0.0, 0.06, 0.54}

\def\nn{\nonumber \\}

\def\lmk{\left(}
\def\rmk{\right)}
\def\lkk{\left[}
\def\rkk{\right]}

\newcommand{\bel}[1] {\begin{equation}\label{#1}}
\newcommand{\beal}[1] {\begin{eqnarray}\label{#1}}
\newcommand{\bea}{\begin{eqnarray}} 
\newcommand{\eea}{\end{eqnarray}}

\usepackage{comment}

\begin{document}

\title{Compact objects as the catalysts for vacuum decays}

\author{Naritaka Oshita$^{1,2}$}
\author{Masaki Yamada$^{3}$}
\author{Masahide Yamaguchi$^{4}$}
\affiliation{
  $^1$Research Center for the Early Universe (RESCEU), Graduate School
  of Science,\\ The University of Tokyo, Tokyo 113-0033, Japan
}
\affiliation{
  $^2$Department of Physics, Graduate School of Science,\\ The University of Tokyo, Tokyo 113-0033, Japan
}
\affiliation{
  $^3$Institute of Cosmology, Department of Physics and Astronomy, Tufts University, Medford,
MA 02155, USA
}
\affiliation{
  $^4$Department of Physics, Tokyo Institute of Technology,
2-12-1 Ookayama, Meguro-ku, Tokyo 152-8551, Japan
}

\date{\today}

\begin{abstract}
We discuss vacuum decays catalyzed by spherical and horizonless objects
and show that an ultra compact object could catalyze a vacuum decay around it within the cosmological time.
The catalytic effect of a horizonless compact object could be more efficient than that of a black hole
since in this case there is no suppression of the decay rate due to the decrement of its
Bekestein entropy.
If there exists another minimum with AdS vacuum in the Higgs potential at a high energy scale, the abundance of compact objects
such as monopoles, neutron stars, axion stars, oscillons, Q-balls, black hole remnants, gravastars and so on, could be severely
constrained. 
We find that an efficient enhancement of nucleation rate occurs 
when the size of the compact object is comparable to its Schwarzschild radius 
and the bubble radius. 
\end{abstract}

\pacs{}

\maketitle


\section{introduction}
Compact objects are ubiquitous in high-energy physics as well as astrophysics 
and play significant roles in cosmological history of the Universe. 
To name a few, 
monopoles~\cite{Vilenkin:2000jqa}, 
Q-balls~\cite{Coleman:1985ki, Kusenko:1997si,Enqvist:1997si,Enqvist:1998en,Kasuya:1999wu,Kasuya:2000wx,Kasuya:2001hg, Lee:1988ag, Kasuya:2015uka, Hong:2015wga}, 
oscillons~\cite{Bogolyubsky:1976nx,Segur:1987mg,Gleiser:1993pt,Copeland:1995fq,Gleiser:1999tj, Kasuya:2002zs,Gleiser:2004an,Fodor:2006zs,Hindmarsh:2006ur, Amin:2011hj, Mukaida:2016hwd}, 
boson stars (including axion stars)\cite{Ruffini:1969qy, Hogan:1988mp,Kolb:1993zz,Seidel:1993zk,Hu:2000ke,Guzman:2006yc,Sikivie:2009qn,Liebling:2012fv,
Guth:2014hsa, Eby:2014fya, Braaten:2015eeu, Braaten:2016kzc, Hui:2016ltb, Eby:2016cnq, Eby:2017teq, Eby:2018ufi}, 
gravastars~\cite{Mazur:2001fv,Mazur:2004fk} (see also \cite{Carballo-Rubio:2017tlh}), 
neutron stars, 
black hole (BH) remnants~\cite{Aharonov:1987tp,Adler:2001vs}, 
and (primordial) BHs~\cite{Hawking:1971ei, Carr:1974nx, Carr:1975qj, Ivanov:1994pa, GarciaBellido:1996qt, Kawasaki:1997ju, Yokoyama:1998pt, Garriga:2015fdk, Garcia-Bellido:2017mdw, Deng:2017uwc, Hertzberg:2017dkh}
are examples that have been studied extensively in the literature for several decades. 
Pursuing consistency of these objects in cosmology and astrophysics 
is important to construct a realistic particle physics model 
and is complementary to high-energy colliders to find a new physics beyond the standard model.

It has been proposed that BHs may be objects catalyzing
vacuum decays around them~\cite{Berezin:1987ea, Arnold:1989cq, Berezin:1990qs, Gomberoff:2003zh, Garriga:2004nm, Gregory:2013hja, Burda:2015isa, Burda:2015yfa, Chen:2017suz, Mukaida:2017bgd, Kohri:2017ybt}, which was pioneered by Hiscock~\cite{Hiscock:1987hn}.
The abundance of the catalyzing objects
should be small enough to avoid the nucleation of AdS vacuum bubble within our observable Universe 
until present. 
Actually, this is particularly important in the standard model of particle physics~\cite{Burda:2015isa,Burda:2016mou,Cole:2017gle}, where 
the Higgs potential could develop a AdS vacuum at a high energy scale 
because of the running of quartic coupling 
\cite{
Sher:1988mj,Arnold:1989cb,Altarelli:1994rb,Espinosa:1995se, 
Casas:1996aq,Hambye:1996wb,Isidori:2001bm,Espinosa:2007qp,Ellis:2009tp,Bezrukov:2012sa,Bednyakov:2015sca,EliasMiro:2011aa,Degrassi:2012ry,Buttazzo:2013uya,Branchina:2013jra,Bednyakov:2015sca}.
According to their result, even a single BH within our observable Universe leads to the bubble nucleation 
if its mass is small enough. 

One may wonder what property of BHs contributes to the
promotion of a vacuum decay around it.
Gregory, Moss, and Withers found \cite{Gregory:2013hja,Burda:2015yfa} that the exponential factor
of a vacuum decay rate around a BH is determined by
two factors, $\Gamma \propto e^{-B + \Delta S}$,
where $\Gamma$ is the vacuum decay rate, $\Delta S$
is the change of Bekenstein entropy of the BH,
and $B$ is an on-shell Euclidean action depending on
the Euclidean dynamics of a bubble wall. Then they found that
the decrement of $B$ due to gravity of a BH overwhelms the entropy decrement.
Although they found an extremely large enhancement of bubble nucleation rate around a BH, 
it has been discussed that the main effect comes from the thermal
fluctuation due to the Hawking radiation~\cite{Mukaida:2017bgd}. This
implies that the nucleation rate is overestimated because the same
effect generates a thermal potential that tends to stabilize the Higgs at the symmetric phase~\cite{Gorbunov:2017fhq, Mukaida:2017bgd, Kohri:2017ybt}
or because the thermal effect of Hawking radiation should be small for a large bubble. 
Thus, though the bubble nucleation rate is still enhanced around a
BH because of the effect of gravity, it is not so large as expected before.
If gravity of a BH mainly contributes to
the promotion of a vacuum decay,
it is meaningful to consider the catalyzing effect even around
horizonless objects. 
The absence of horizons is equivalent to the
absence of the suppression factor due to the change of Bekenstein entropy $e^{\Delta S}$, 
and therefore, horizonless compact objects may be
more important candidates of catalyzing objects for
vacuum decays.
In this manuscript, we discuss such a vacuum decay around
a spherical horizonless object as a catalyzing one.

This paper is organized as follows. 
In Sec.~\ref{sub:form}, we explain the formalism to calculate 
the bubble nucleation rate around a generic compact object. 
We use a Gaussian density function for the object 
as an example to calculate the nucleation rate in Sec.~\ref{sub:Gaussian}. 
We will see that the efficient enhancement occurs 
if the radius of the compact object, its Schwarzschild radius, 
and the radius of the nucleated bubble are of the same order with each other. 
We then discuss a parameter region for a t'Hooft-Polyakov monopole 
that is excluded because of the nucleation of AdS vacuum in Sec.~\ref{sub:monopole}. 
In Sec.~\ref{sec:comparison}, we discuss differences from the bubble nucleation around a black hole. 
We will see that the nucleation rate is more enhanced around a horizonless compact object 
than around a black hole with the same total mass. 
Our conclusions are summarized in Sec.~\ref{sec:conclusions}.  

\section{bubble nucleation around a compact object}
\label{sec:bubble_nuc}

\subsection{Formalism}
\label{sub:form}
We consider a nucleation of a thin wall vacuum bubble 
around a spherical object. 
If we assume that the system is static, 
the metric inside and outside of the bubble 
can be written as 
\beq
ds^2 = - C_{\pm} (r_{\pm}) dt_{\pm}^2 + D_{\pm} (r_{\pm}) dr_{\pm}^2 +r_{\pm}^2 d\Omega_{2}^2,
\label{metric1}
\eeq
where $C_\pm$ and $D_\pm$ are determined by the Einstein equation and will be specified later. 
The quantities associated with the outer
and inner region are labeled by
the suffix ``$+$" and by ``$-$", respectively.

\begin{figure}[t]
  \begin{center}
    \includegraphics[keepaspectratio=true,height=40mm]{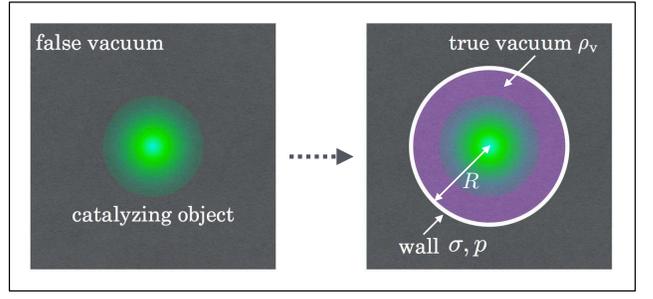}
  \end{center}
  \caption{
  A schematic picture showing a vacuum decay catalyzed by a static and spherical object. 
}%
  \label{fig1}
\end{figure}

The thin wall vacuum bubble can be characterized by its energy density, $\sigma$,
and pressure, $p$. 
The ratio of $p$ to $\sigma$, $w \equiv p/\sigma$
(equation-of-state parameter), is assumed to be a constant. 
We here choose the scale of radial coordinates $r_{\pm}$ so that $r_+ = r_- \equiv R$ on the wall.
A schematic picture showing the vacuum decay process we here assume is depicted in Fig. \ref{fig1}.

Introducing the extrinsic curvature on the outer (inner) surface of the wall, $K^{(+)}_{AB}$ ($K^{(-)}_{AB}$),
the energy-momentum tensor (EMT) of wall, $S_{AB}$, and the induced metric on the wall, $h_{AB}$,
the dynamics of the thin wall with $\xi^A=(\tau,\theta,\phi)$ is described by the Israel junction conditions as
\begin{align}
&K^{(+)}_{AB} - K^{(-)}_{AB} = -8  \pi G \left( S_{AB}- \frac{1}{2}
 h_{AB} S \right), \label{israel}\\
&\sqrt{C_{\pm}D_{\pm}} K^{(\pm)}_{AB} = \text{diag} \left( - \frac{d
 \beta_{\pm}}{dR}, \beta_{\pm} R, \beta_{\pm} R \sin^2 \theta \right),\\
&S^A{}_B \equiv \text{diag} \left( -\sigma, p, p \right), h_{AB}
 \equiv \text{diag} \left( -1, R^2, R^2 \sin^2\theta \right),
\end{align}
where 
\beq
 \beta_{\pm} \equiv \epsilon_{\pm} \sqrt{C_{\pm} + C_{\pm}D_{\pm} (dR/d\tau)^2}, 
\eeq
and $\tau$ is the proper time of the wall
and $\epsilon_{\pm}$ is the sign of spatial components of extrinsic curvature.
We here simply neglect the interaction between the horizonless object
and the bubble except for their gravitational interaction. 
The case with such an interaction being taken into account will be discussed  elsewhere (see also Refs.~\cite{Kusenko:1997hj, Metaxas:2000qf, Pearce:2012jp} in the context of Q-ball in supersymmetric models without taking gravity effects into account). 

We are interested in the decay of Higgs vacuum, 
where the metastable vacuum has a negligibly small vacuum energy 
and the true vacuum has a negative vacuum energy $\rho_{\rm v} < 0$. 
We also introduce a compact object at the origin of the spatial coordinate, 
which modifies the metric because of the nonzero mass density $\rho_{\rm c} (r)$.
For simplicity, we here assume the EMT of the object which gives
the following static solutions of the Einstein equation:
\beq
 C_{\pm} = D_{\pm}^{-1} = f_\pm (r_\pm) \equiv 1-2GM_{\pm}(r_{\pm}) /r_{\pm} + H_{\pm}^2 r_{\pm}^2,
\eeq
with
\bea
&&H_+ = 0, \ H_-^2 \equiv - \frac{8\pi G}{3} \rho_{\text{v}}, 
\\
&&M_{\pm} (r_{\pm}) \equiv \int_0^{r_{\pm}} d\bar{r}_{\pm} 4 \pi \bar{r}_{\pm}^2 \rho_{\text{c}}^{(\pm)} (\bar{r}_{\pm}). 
\eea
If we use an arbitrary mass density, $\rho_{\rm c}^{(\pm)} (r)$, 
the compact object does not satisfy the static Einstein equation unless an appropriate EMT for the chosen $\rho_{\rm c}^{(\pm)} (r)$ exists.
Although in this case the metric (\ref{metric1}) cannot be used,
we expect that we can use it to capture a qualitative result. 
To be more rigorous, in the Appendix,
we calculate the vacuum
decay rate around a gravastar-like object, 
which is constructed to be (approximately) static. 
We specify its interior EMT and use the metrics consistent
with the specified EMT. 
Then one could find that the aforementioned assumption
for the metrics, (\ref{metric1}), does not qualitatively change
our results and main conclusions.

Equation (\ref{israel}) now reduces to the following equations
\begin{align}
& \frac{d}{dR} \left( \beta_- - \beta_+ \right) = - 8\pi G \left( \sigma/2 + p \right), \label{052002}\\
& \lmk \beta_- - \beta_+ \rmk = 4 \pi G \sigma (R) R. \label{052001}
\end{align}
One obtains $\sigma = m^{1-2w} R^{-2 (1+w)}$ by solving (\ref{052002}),
where $m$ is the typical energy scale of the wall, and we can rewrite (\ref{052001}) as
\begin{align}
&\left( \frac{dz}{d \tau'} \right)^2 + V(z) = -1,
\label{EoM}\\
\begin{split}
V(z) &\equiv - \frac{a_+}{z} - \frac{z^2}{4} \left[ \frac{\Delta a \bar{m}^{2w-1} z^{2w-1}}{4 \pi \bar{H}^{2w+1}} \right. \\
&\left. + \frac{z^{2(1+w)} \bar{m}^{2w-1}}{4 \pi \bar{H}^{2w +1}} - 
\frac{4 \pi \bar{H}^{2w +1}}{z^{2(1+w)} \bar{m}^{2w-1}} \right]^2 \leq 0,
\label{potV}
\end{split}
\end{align}
where we re-defined the following non-dimensional variables and parameters:
\begin{align}
\begin{split}
&z \equiv H_- R, \ \tau' \equiv H_- \tau, \ a_{\pm} \equiv 2 G M_{\pm} H_-,\\
&\Delta a \equiv 2G (M_+ - M_-) H_-, \ \bar{m} \equiv m/M_{\text{Pl}}, \ \bar{H} \equiv H_-/M_{\text{Pl}}.
\end{split}
\end{align}
The parameters $\bar{m}$ and $\bar{H}$ are the ones in the Planck units $M_{\text{Pl}}=1/\sqrt{G}$. 
Implementing the Wick rotation, $\tau = -i \tau_{\text{E}}$, (\ref{EoM}) gives the bounce solution that describes the bubble nucleation process.
In the following, we assume that matter fields forming a compact object has no interaction with another matter field which eventually undergoes the phase transition. In this case, the transition would not change the mass of the object, and therefore, $\Delta a = 0$, that is, $M_+ (R) = M_- (R) \equiv M (R)$, is valid in (\ref{potV}). Even if this is not the case, $\Delta a = 0$ is a good approximation, provided that $|M_+ - M_-| \ll (4 \pi /3)R^3 |\rho_{\text{v}}|$, for which the first term in the square-bracket in (\ref{potV}) is negligible compared to the second term.

The Euclidean action, $B_{\text{co}}$, can be calculated from the bounce solution with the following integration \cite{Gregory:2013hja}:
\bea
B_{\text{co}} = \frac{1}{4 G} \int d \tau_{\text{E}} (2R - 6 G M + 2GM'R) \left( \frac{\beta_+}{f_+} - \frac{\beta_-}{f_-} \right).
\nn \label{BCO1}
\eea
The transition rate, $\Gamma_{\text{D}}$, can be estimated as
\begin{equation}
\Gamma_{\text{D}} \sim R_{\text{CDL}}^{-1} \sqrt{\frac{B_{\text{co}}}{2 \pi}} \exp({-B_{\text{co}}}),
\label{BCO2}
\end{equation}
where we estimate the prefactor by taking a factor of $\sqrt{B_{\text{co}}/2 \pi}$ for the zero mode associated with the time-translation of the instanton
and we use the light crossing time of the bubble, $R_{\text{CDL}}$, as a
rough estimate of the determinant of fluctuations, which will be
defined more precisely below.

\subsection{Results for Gaussian mass function}
\label{sub:Gaussian}
As an example, we consider the case where the density distribution of the horizonless object is given by the Gaussian form: 
\beq
 \rho_{\text{c}} (r) = \rho_0 e^{- r^2/\xi^2},
\eeq
where $\rho_0$ and $\xi$ represent the typical mass density and the size of the compact object, respectively, and $\rho_{\text{c}} (R) \equiv \rho_{\text{c}}^{(+)} (R) = \rho_{\text{c}}^{(-)} (R)$. 
Motivated by the Higgs vacuum decay, we take $\bar{H} = 10^{-6}$, $\bar{m} = 6 \times 10^{-4}$, and $w= -1$ 
throughout this manuscript\footnote{We assume that the absolute value of true-vacuum energy density, $\rho_{\text{v}}$, and the the height of the Higgs potential barrier, $V_{\text{max}}$, are of the order of the GUT scale $ \sim 10^{-12} M_{\text{Pl}}^4$. The vacuum expectation value of the true-vacuum state, $\phi_0$, is assumed to be $\phi_0 \sim 10^{-3} M_{Pl}$. This gives $\bar{H} \sim 10^{-6} M_{\text{Pl}}$ and $\bar{m} = (\sigma/M_{\text{Pl}}^3)^{1/3} \sim \left(\kappa \sqrt{V_{\text{max}}/M_{\text{Pl}}^4} \times \phi_0/M_{\text{Pl}} \right)^{1/3} \sim 6 \times 10^{-4}$, where the constant $\kappa$ depends on the details of the Higgs potential \cite{Burda:2015yfa} and we here take $\kappa \sim 0.1$.}.
Here, we implicitly assume that the Higgs potential is supplemented by a non-renormalizable $\phi^6$ term as considered in Ref.~\cite{Burda:2015yfa} 
so that we can use the thin-wall approximation. 
We also take $\xi = 10^{3} M_{\text{Pl}}^{-1}$ as an example.

\begin{figure}[t]
  \begin{center}
    \includegraphics[keepaspectratio=true,height=95mm]{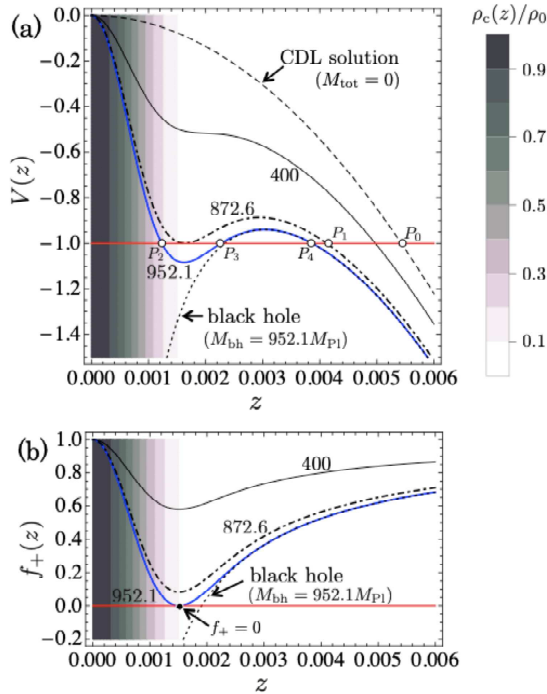}
  \end{center}
  \caption{
The effective potential (a) and $f_{+}$ (b) for the horizonless object with $M_{\text{tot}}/M_{\text{Pl}} = 0$ (CDL solution),
$400$ (a black solid line), $872.6$ (a black dashed-dotted line),
and $952.1$ (a blue solid line)
and for a BH with $M_{\text{tot}} / M_{\text{Pl}}= 952.1$ (a black dotted line) are shown.
}%
  \label{fig2}
\end{figure}

Effective potentials governing the wall position ($V(z)$) for the above parameters are plotted in Fig. \ref{fig2}-(a). 
The dashed line represents the case of Coleman De-Luccia (CDL) tunneling, 
where $M_{\text{tot}}/M_{\text{Pl}} = 0$ with $M_{\text{tot}} \equiv \int^{\infty}_{0} dr' 4 \pi r' {}^2 \rho_{\text{c}} (r')$. In this case, a bubble is nucleated at 
the point $P_0$ i.e. with the radius $R \simeq \alpha H_-^{-1} \equiv R_{\text{CDL}}$ for $\alpha \equiv 8 \pi Gm^3/H_- \ll 1$ \cite{Coleman:1980aw,Gregory:2013hja}. 
As we increase $M_{\text{tot}}/M_{\text{Pl}}$, 
the effective potential becomes lower. 
We plot the cases of $M_{\text{tot}}/M_{\text{Pl}} = 400$ (a black solid line), $872.6$ (a black dashed-dotted line),
and $952.1$ (a blue solid line). 

A nucleated vacuum bubble with $0 \leq M_{\text{tot}} / M_{\text{Pl}} \lesssim 872.6$ initially has its wall radius between $P_1$ and $P_0$
(black open circles in Fig. \ref{fig2}) and would expand soon after its nucleation. 
A nucleated bubble around the horizonless object with
$872.6 \lesssim M_{\text{tot}} / M_{\text{Pl}} \lesssim 952.1$ would be trapped
between $P_2$ and $P_3$, where the gravitational force and bubble tension are balanced,
and then, it may eventually tunnel to a larger bubble, whose wall is in between $P_4$ and $P_1$.
If the mass is larger than or equal to $952.1 M_{\text{Pl}}$, one has $f_{+} = 0$ (a black filled circle in Fig. \ref{fig2}-(b)), that is, a BH forms. 

One finds that the effective potential can be drastically distorted because of the gravitational effect of the horizonless object,
which makes a bubble wall nucleated around a catalyzing object smaller compared to a CDL bubble ($P_0$ in Fig. \ref{fig2}-(a)). 
The distortion of the potential largely enhances the nucleation rate of vacuum bubble and the nucleation of bubbles could occur within the cosmological time
as will be shown in the following.

\begin{figure}[t]
  \begin{center}
    \includegraphics[keepaspectratio=true,height=60mm]{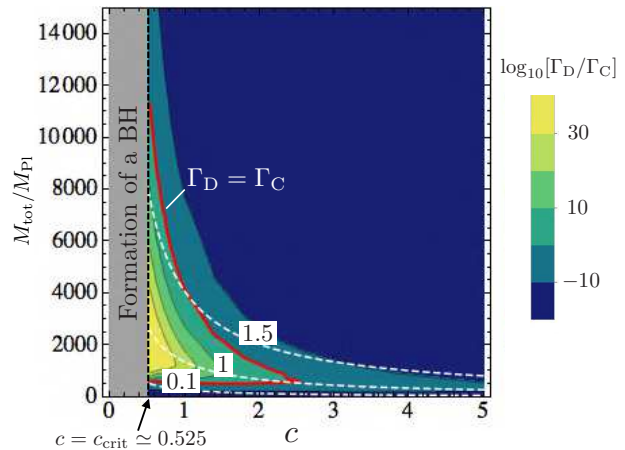}
  \end{center}
  \caption{
A plot of the ratio of the decay rate, $\Gamma_{\text{D}}$, to the inverse of the cosmological time, $\Gamma_{\text{C}}$,
as a function of the mass and compactness of the horizonless object. 
The contour of $\Gamma_{\text{D}} = \Gamma_{\text{C}}$ (red solid line) and contours of $\xi/R_{\text{CDL}}$ (white dashed lines)
are marked for reference.
In the case of $c \leq c_{\text{crit}} \simeq 0.525$ (gray shaded region),
the object inevitably collapses to a BH since a function $f_+ (r)$ has zero points there.
}%
  \label{fig3}
\end{figure}

In Fig. \ref{fig3}, the ratio of the vacuum decay rate, $\Gamma_{\text{D}}$, to the inverse of the cosmological time, $\Gamma_{\text{C}} \equiv H_{\text{C}} \simeq 10^{-61}
M_{\text{Pl}}$, is shown in the range of $1 \leq M_{\text{tot}}/ M_{\text{Pl}} \leq 15000$ and of $c \leq 5$,
where we define the compactness parameter\footnote{Although there is an ambiguity in the definition of the radius because of the thick boundary of the Gaussian mass distribution, this ambiguity just changes the scale of $c$-axis in Fig. \ref{fig3} and does not affect the result shown there.} as 
\beq
 c \equiv \xi / (2GM_{\text{tot}}). 
\eeq
In our setup, we find that the existence of even a single horizonless object 
with $M_{\text{tot}}/ M_{\text{Pl}}$ and $c$ 
within the region enclosed by the red line in the figure 
(i.e., $10^3 \lesssim M_{\text{tot}}/ M_{\text{Pl}} \lesssim 10^4$ and with $c \lesssim 2$) 
would be excluded since a bubble would be nucleated around it within the cosmological time. 

We show the contours of $\xi/R_{\text{CDL}}$ 
as white dashed lines in Fig. \ref{fig3}. 
They indicate that an efficient enhancement occurs 
only when 
the radius of the nucleated bubble (which is of the same order with the CDL radius) is comparable to 
that of the compact object. 
An efficient enhancement also requires a small compactness 
so that the gravity effect is efficient around the dense compact object. 
Therefore, we conclude that 
the bubble nucleation rate is drastically enhanced around a compact object 
if the size of the horizonless object is comparable with the radius of CDL bubble 
and its compactness is of the order of unity. 

\subsection{Constraint on the abundance of compact objects}
\label{sub:monopole}
The nucleation of the anti-de Sitter vacuum bubble, whose origin could be the Higgs instability, within the cosmological time is obviously conflict with the present Universe not filled by the negative vacuum energy. 
It would expand with speeds comparable with the light speed soon after its nucleation, which would lead to the Universe filled by the negative vacuum energy.
Since a compact object could be a catalyst for the vacuum decay, we can put the constraint on the abundance of horizonless objects in the Universe.

For instance, depending on parameters characterizing magnetic monopoles,
the monopoles could be ultra compact. 
Suppose that there is a (hidden) non-Abelian gauge field that is spontaneously broken 
by a (hidden) Higgs field. If the vacuum manifold has 
a non-trivial second homotopy group, monopoles arise at the spontaneously symmetry breaking. 
Introducing the vacuum expectation value (VEV) of the hidden Higgs field, $v$,
the mass and size of a t'Hooft-Polyakov monopole, denoted by $M_{\text{mono}}$ and $R_{\text{mono}}$,
respectively, can be estimated as 
\bea
 &&M_{\text{mono}} \sim v/\sqrt{\alpha_{\text{G}}} 
 \\
 &&R_{\text{mono}} \sim (\sqrt{\alpha_{\text{G}}} v)^{-1}, 
\eea
where 
$\alpha_{\text{G}}$ is the running gauge coupling constant for the non-Abelian gauge interaction.
Imposing the ultra compact condition,
$c \simeq R_{\text{mono}}/ (2G M_{ \text{mono}}) \sim 1$,
one obtains $v \sim M_{\text{Pl}}$ and $R_{\text{mono}} \sim \ell_{\text{Pl}}/\sqrt{\alpha_{\text{G}}}$.
Therefore, as long as the Higgs potential accommodates
a second lower minimum due to the Higgs instability,
parameter regions which realize $v \sim M_{\text{Pl}}$ and
$R_{\text{CDL}} \sim R_{\text{mono}} \sim \ell_{\text{Pl}} / \sqrt{\alpha_{\text{G}}}$ should be
excluded in order to be consistent with the present Universe
not filled by the anti-de Sitter vacuum. 
Since $R_{\rm CDL} = 8 \pi G m^3 / H_-^2 \simeq 5 \times 10^3$, 
$\alpha_{\rm G}$ should be as small as $3 \times 10^{-8}$ to nucleate the Higgs vacuum bubble.

\section{Comparison with the catalyzing effect of black holes}
\label{sec:comparison}
Now we compare our results with the case of bubble nucleation around a BH, 
which has been extensively discussed in the literature. 
In Ref.~\cite{Gregory:2013hja}, Gregory, Moss, and Withers pointed out that
the Bekenstein entropy of a BH with mass $M_{\text{tot}}$ may contribute to the vacuum decay rates as 
\begin{align}
\Gamma_{\text{D}} &\sim R_{\text{CDL}}^{-1} \sqrt{\frac{I_{\text{E}}}{2 \pi}} e^{-I_{\text{E}}} = R_{\text{CDL}}^{-1} \sqrt{\frac{I_{\text{E}}}{2 \pi}} e^{-B_{\text{bh}} + \Delta S},
\label{GAD}\\
B_{\text{bh}} & \equiv \frac{1}{4 G} \oint d \tau_{\text{E}} (2R - 6 G M_{\text{tot}}) \left( \frac{\beta_+}{f_+} - \frac{\beta_-}{f_-} \right),
\label{GRE}
\end{align}
where $I_{\text{E}}$ is the total Euclidean action 
and $B_{\text{bh}}$ is the bulk component of the on-shell Euclidean action
depending on the Euclidean dynamics of a vacuum bubble. 
Contributions from the conical singularities on the Euclidean
manifolds before and after the vacuum decay 
lead to a factor of 
$\Delta S$, which is equivalent to the change of the Bekenstein entropy
of a catalyzing BH. 

Note that even horizonless compact objects can emit Hawking
radiation (see, e.g., Refs. \cite{H.Kawai2013a, H.Kawai2015b,
H.Kawai2017c}) because of the vacuum polarization in a strong
gravitational field and its thermal effect on the Higgs potential may
have to be taken into account.  The details of the Higgs potential are
characterized by the parameters (i.e. $\bar{H}$, $\bar{m}$, and $w$) in
the thin-wall approximation, and therefore, those parameters could be
affected by such a thermal effect in our setup\footnote{Although it might be possible that the BH
mass changes due to the bubble nucleation~\cite{Gregory:2013hja,
Burda:2015isa, Burda:2015yfa, Burda:2016mou}, it was argued that it
could be closely related to the thermal excitation of bubble due to the
Hawking radiation~\cite{Gorbunov:2017fhq, Mukaida:2017bgd}.}.

When a BH efficiently catalyzes the vacuum decay, the size of the BH, $2GM_{\text{tot}}$, is comparable with the CDL bubble radius, $R_{\text{CDL}}$, and
the prefactor in (\ref{GAD}) can be rewritten as $(GM_{\text{tot}})^{-1} \sqrt{I_{\text{E}}/2 \pi}$, which is consistent with the prefactor in Ref. \cite{Gregory:2013hja}.
The Bekenstein entropy decreases because of the vacuum decay since the decrease of the vacuum
energy surrounding the BH makes the area of its event horizon smaller. 
Although the gravity effect is strong around a BH, 
the vacuum decay rate would be suppressed by the change of Bekenstein entropy:
\beq
 \Delta S = \pi \lkk R_{\rm h,-}^2 - (2 G M_{\rm tot})^2 \rkk < 0, 
\eeq
where the horizon radius $R_{\rm h,-}$ after the bubble nucleation is defined by $f_- (R_{\rm h,-}) = 0$. 
\begin{figure}[t]
  \begin{center}
    \includegraphics[keepaspectratio=true,height=88mm]{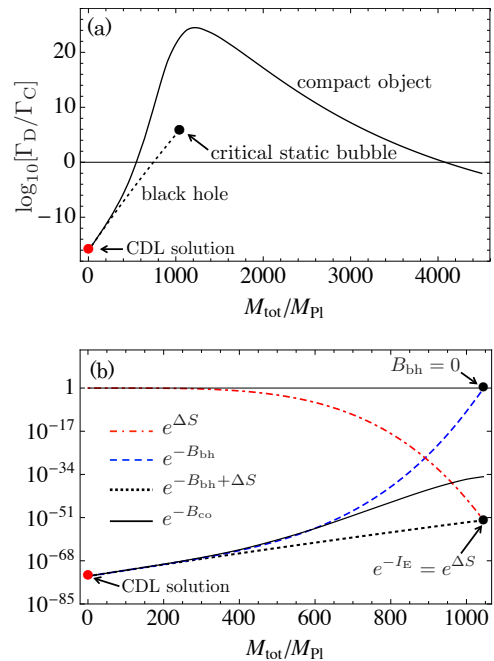}
  \end{center}
  \caption{
The vacuum decay rates around a BH with mass $M_{\text{tot}}$ (a dotted line)
and that around a horizonless compact object, whose mass is $M_{\text{tot}}$ and compactness
is fixed with $\xi/2GM_{\text{tot}} = 1$,
(a solid line) are shown. Red and black points show the decay rate of the CDL solution
and a critical static solution, respectively.
}%
  \label{fig4}
\end{figure}

Horizonless compact objects have no Bekenstein entropy, 
so that they could more efficiently catalyze vacuum decays than BHs do. 
We here compare the vacuum decay rate around the horizonless compact object, whose mass is $M_{\text{tot}}$ and compactness is fixed with
$c = 1$, with that around a BH whose mass is $M_{\text{tot}}$. 
The result is shown in Fig. \ref{fig4}.
Both the BH and horizonless compact object efficiently catalyze the vacuum decay compared to the CDL solution shown as red points. 
If there were no contribution of Bekenstein entropy on the decay rate around the BH, the exponential
factor for the BH would be larger than that for the horizonless object (blue dashed line in Fig. \ref{fig4}-(b)).
However, 
the decay rate with the compact object (black solid line) is larger than that with the BH (black dotted line) thanks to the absence of the decrement of Bekenstein entropy (red dashed-dotted line).

A bubble nucleated around a BH with $M_{\text{tot}} = M_{\text{crit}} \simeq 1045 M_{\text{Pl}}$
is static (black points in Fig. \ref{fig4}) because of the perfect balance between the gravity of the BH
and bubble's tension.
On the other hand, there is no well-defined Euclidean solution for a BH with $M_{\text{tot}} \geq M_{\text{crit}}$ \cite{Gregory:2013hja}.

\section{Conclusions}
\label{sec:conclusions}
We have discussed a role of a horizonless compact object as a catalyst for a vacuum decay.
As long as the interaction between a bubble and a catalyzing object is
negligible, our results do not depend on the details of the object much
and its gravity plays an essential role in the catalyzing process.
The universality of our result is also discussed in the Appendix.
This suggests that one can put some constraints on the abundance of various kinds of horizonless compact objects, such as monopoles,
Q-balls, Boson stars, gravastars, BH remnants, and so on. 
In particular, the Higgs vacuum may decay into an AdS vacuum 
if there exists even a single compact object 
whose radius is comparable to its Schwarzschild radius 
and the CDL bubble radius. 
For instance, depending on parameters characterizing magnetic monopoles,
the monopoles could be ultra compact.
As long as the Higgs potential accommodates
a second lower minimum due to the Higgs instability,
parameter regions which realize the Planck-scale GUT symmetry breaking with $R_{\text{CDL}} \sim R_{\text{mono}}$ should be
excluded in order to be consistent with the present Universe
not filled by the anti-de Sitter vacuum. 
More realistic cases may arise for Boson stars, oscillons, and Q-balls. 
In addition, in case that a single compact object is not enough to
catalyze the Higgs vacuum to decay into the AdS one, multiple ones
could do it, which leads to new constraints on the abundance of such
a compact object.

It is also interesting to note that the catalyzing effect of horizonless
objects is more efficient compared to that of BHs
since there is no suppression
of vacuum decay rate due to the decrement of Bekenstein entropies.
Therefore, if there had been some ultra compact objects in the Universe,
they could have played a critical role in the cosmological sense.

Finally, we comment on the case where the compact object has an interaction with the nucleated bubble, namely, the Higgs field. 
In this case, the mass of the compact object can change due to the bubble nucleation. 
Because of the conservation of energy, 
the nucleated bubble can use the mass difference of the compact object 
and the nucleation rate can be drastically enhanced. 
This is similar to the case of bubble nucleation in a finite temperature plasma, 
where a bubble can use the thermal energy to be excited with a finite energy. 
This is also similar to the case for a bubble nucleation around a black hole 
with the thermal effect of Hawking radiation, where the mass of black hole changes after the bubble nucleation. 
However, as the thermal effect stabilize the Higgs potential to the symmetric phase in these cases, 
the interaction between the compact object and the Higgs field may lead to an effective potential that stabilize the Higgs potential. 
Still, this results in a more efficient enhancement for the nucleation rate 
and is an interesting possibility for many particle physics models.

\acknowledgements
Masaki Yamada thanks F. Takahashi for a stimulating discussion. 
This work was supported by Grant-in-Aid for JSPS Fellow No. 16J01780
(N.O.), JP25287054 (M.Y.), JP15H05888 (M.Y.), JP18H04579 (M.Y.), and
JP18K18764 (M.Y.).

\appendix

\section{Static gravastar-like objects}

Throughout the main text we assume that 
the compact object is static at least during the nucleation process 
and the metric is given by the static solution (\ref{metric1}). 
This is (approximately) justified 
for most of the realistic situations, like neutron stars, 
boson stars, oscillons, monopoles, and Q-balls, and so on. 
However, 
the density function $\rho_{\rm c} (r)$ as well as the metric functions $C_{\pm}$ and $D_{\pm}$ should be carefully 
chosen so that it is a static solution to the Einstein equation. 
In this Appendix, we consider a gravastar-like object 
to show that the result in Fig. \ref{fig3} does not change qualitatively 
as long as we choose those functions carefully to (approximately) satisfy the static equilibrium.

We use the following EMT for the gravastar-like object: 
\begin{align}
T^{\mu}_{\nu} &= \text{diag} (- \rho (r), p (r), p (r), p (r)),\\
\rho (r) &\equiv \rho_{0} \frac{1-\tanh{\left({(r-\xi)/\delta}\right)}}{2}
+\rho_{\text{v}} = -p (r)
\end{align}
with $r < R$, where $T^{\mu}_{\nu}$ is the bubble interior EMT
and $\delta$ represents the thickness of the boundary of
the gravastar-like object.
When $\delta \ll \xi$, one can use the thin wall approximation
and the bubble interior energy density, $\rho$, is written as
\begin{equation}
\rho (r) \simeq
\begin{cases}
\rho_0 + \rho_{\text{v}} \equiv \rho_{\text{in}} > 0 & \xi > r \\
\rho_{\text{v}} < 0 & \xi < r < R,
\end{cases}
\end{equation}
where the energy density of the gravastar-like object $\rho_0$ is constant.
Assuming the form of its pressure as $p = - \rho$,
the inner metric of the gravastar-like object is given by
\begin{align}
g_{\mu \nu}^{(\text{in})} &= \text{diag} (-f_{\text{in}} (r_{\text{in}}),
f_{\text{in}}^{-1} (r_{\text{in}}),
r_{\text{in}}^2, r_{\text{in}}^2\sin^2{\theta}),\\
f_{\text{in}} (r) &\equiv 1-H_{\text{in}}^2 r^2,
\end{align}
where $H_{\text{in}}^2 \equiv (8 \pi G/3) \rho_{\text{in}}$ and $r_{\text{in}}$ is the radial coordinate inside the object and
we set its scale so that $r_{\text{in}} = r_- = \xi$ on the boundary of
gravastar-like object.

Although the bulk of gravastar-like object has its static metric,
whether or not its boundary is also static should be determined by the Israel
junction condition that is available only when the thickness of its boundary is smaller than its radius, $\delta \ll \xi$.
In the thin wall approximation, the boundary can be characterized
only by its energy density, $\sigma_{c}$, and pressure, $p_{c}$.
Introducing the equation-of-state parameter, $w_c \equiv p_c/\sigma_c$,
one has the Israel junction conditions:
\begin{align}
\beta_{\text{in}} - \beta_- &= 4 \pi G \sigma_c (\xi) \xi,
\label{appththcom}\\
\frac{d}{d \xi} \left( \beta_{\text{in}} - \beta_- \right)
&= - 8 \pi G \sigma_c (\xi) \left( 1/2 + w_c \right),
\label{appttcom}
\end{align}
where $\beta_{\text{in}} \equiv \epsilon_{\text{in}} \sqrt{f_{\text{in}} (\xi) +
(d\xi/d\tau_c)^2}$ and $\tau_c$ is the proper time on the boundary
of the object.
\begin{figure}[b]
  \begin{center}
    \includegraphics[keepaspectratio=true,height=45mm]{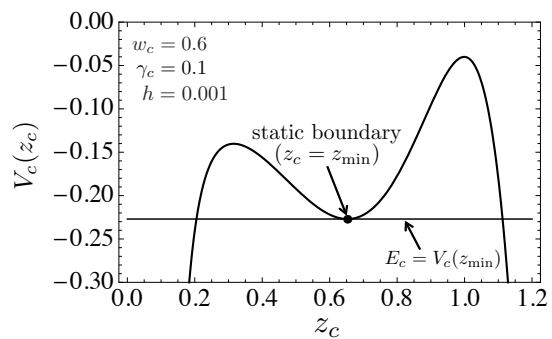}
  \end{center}
  \caption{
  A plot of the effective potential, $V_{c} (z_c)$, with $w_c=0.6$, $\gamma_c = 0.1$, and $h=10^{-3}$.
}%
  \label{fig_app1}
\end{figure}
Solving (\ref{appththcom}) and (\ref{appttcom}),
one has the form of $\sigma_c (\xi) \equiv m_c^{-1-2 w_c} \xi^{-2 (1+w_c)}$.
Substituting $\sigma_c (\xi)$ into (\ref{appththcom}), one has
\begin{align}
&\left( \frac{dz_c}{d\tau'_c} \right)^2 + V_c (z_c) = E_c,\\
&V_c(z_c) \equiv - \frac{4 \gamma_c^2}{1+h^2} z_c^2 - z_c^{4w_c} \left( 1-z_c^3 + \frac{\gamma_c^2}{z_c^{1+4w_c}} \right)^2,
\label{app_vc}
\end{align}
where we defined the following non-dimensional variables and parameters:
\begin{align}
h & \equiv H_- / H_\text{in},\\
z_c^3 &\equiv \left( \frac{1+h^2}{2GM_{\text{tot}} H_-} \right) H_-^3 \xi^3, \label{app_z} \\
\tau'_c &\equiv \frac{\sqrt{1+h^2}}{2 \gamma_c} H_- \tau_c, \\
\gamma_c^2 &\equiv H_-^{2 (1+4 w_c)/3}
\frac{(4 \pi G m^{-1-2w_c})^2}{2GM_{\text{tot}}}
\left( \frac{1+h^2}{2GM_{\text{tot}}} \right)^{(1+4w_c)/3},\\
E_c & \equiv - \frac{4 \gamma_c^2}{(2GM_{\text{tot}} H_-)^{2/3} (1+h^2)^{1/3}}. \label{app_Eform}
\end{align}

Now one obtains stable solutions by appropriately choosing the parameters.
An effective potential, $V_c (z_c)$, governing the position of the boundary is plotted in Fig \ref{fig_app1}.
One finds a stable and static solution (a black filled circle in Fig. \ref{fig_app1}),
at which its radius is $z_c = z_{\text{min}}$ and $E_c = V_c(z_{\text{min}})$.
An effective potential governing the dynamics of the boundary before
the phase transition is obtained just by taking $h = 0$ in (\ref{app_vc}).
Therefore, the effective potential, $V_c (z_c)$, is almost not affected by
the phase transition as long as $h \ll 1$ is hold (see (\ref{app_vc})).
In this case, 
the gravastar-like object remains almost static 
even after the bubble nucleation 
and we can safely use the static metric (\ref{metric1}) to calculate the bubble nucleation rate.

Fixing $h (\ll 1)$, $\gamma_c$, and $w_c$, one may obtain a static solution,
$dV_c(z_c=z_{\text{min}})/dz_c = 0$, and the total mass and size of the gravastar-like object are given by
\begin{align}
M_{\text{tot}} &= \frac{8 \gamma_c^3}{2GH_- (-V_c(z_c=z_{\text{min}}))^{3/2} (1+h^2)^{1/2}},
\label{app_mtot}\\
\xi &= \frac{z_{\text{min}}}{H_-} \left( \frac{2GM_{\text{tot}} H_-}{1+h^2} \right)^{1/3},
\label{app_xi}
\end{align}
where we used (\ref{app_z}) and (\ref{app_Eform}).

\section{Vacuum decay rate around the gravastar-like object}
\begin{figure}[H]
  \begin{center}
    \includegraphics[keepaspectratio=true,height=60mm]{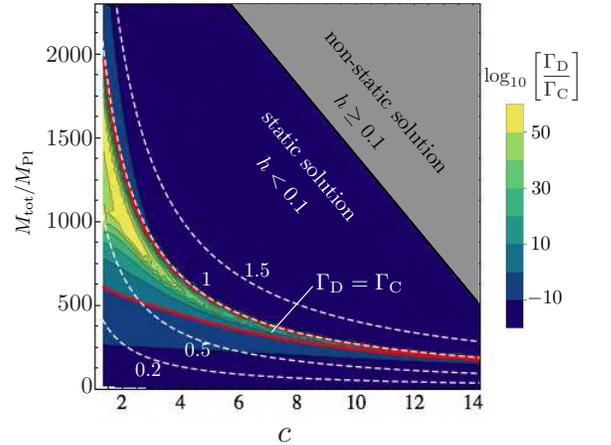}
  \end{center}
  \caption{
 A plot of the ratio $\Gamma_{\text{D}} / \Gamma_{\text{C}}$
as a function of the mass and compactness of the gravastar-like object with
$\bar{H} = 10^{-6}$, $\delta = 0.01 \xi$, $\bar{m} = 6 \times 10^{-4}$,
and $w = -1$.
We here only take into account a parameter region corresponding to
$h \leq 0.1$.
The contours of $\xi/R_{\text{CDL}}$ (white dashed lines) are marked for reference.
}%
  \label{fig_app2}
\end{figure}

Here we calculate the on-shell Euclidean action as a function of $(M_{\text{tot}}, c \equiv \xi/ 2GM_{\text{tot}})$.
Note that we do not take into account a parameter region where $h \geq 0.1$
to approximately keep the gravastar-like object static before and after the
phase transition. The mass function, $M(r)$, in (\ref{BCO1}) should have
the form of
\begin{equation}
M (r) = \int^{r}_{0} dr' 4 \pi r' {}^2 \rho_c (r') \simeq
\begin{cases}
(4 \pi /3) r^3 \rho_0 & \xi > r\\
(4 \pi /3) \xi^3 \rho_0 = M_{\text{tot}} & \xi < r,
\end{cases}
\end{equation}
where $\delta \ll \xi$ is hold.
This gives the metric on the inner and outer surface of the wall:
\begin{equation}
g_{\mu \nu}^{(\pm)} = \text{diag} (-f_{\pm} (R), f_{\pm}^{-1} (R),R^2,R^2\sin^2\theta),
\end{equation}
with
\begin{align}
f_{+} &\simeq
\begin{cases} \displaystyle
1-\frac{2GM_{\text{tot}}}{R} & R > \xi \\
1-H_c^2 R^2 & R < \xi,\\
\end{cases}\label{app_fp}\\
f_{-} &\simeq
\begin{cases} \displaystyle
1 - \frac{2GM_{\text{tot}}}{R} + H_-^2 R^2 & R > \xi \\
1-H_{\text{in}}^2 R^2 & R < \xi.
\end{cases}\label{app_fm}
\end{align}
From (\ref{BCO1}), (\ref{BCO2}), (\ref{app_fp}), and (\ref{app_fm}),
one can calculate the vacuum decay rate. 
Figure \ref{fig_app2} shows the result of $\Gamma_{\text{D}} / \Gamma_{\text{C}}$.
One finds that the result shown in Fig. \ref{fig_app2} is qualitatively
consistent with our conclusion based on the result in Fig. \ref{fig3}.
However, the range of values of compactness, $c$, in which
$\Gamma_{\text{D}} > \Gamma_{\text{C}}$ is satisfied, seems to be sensitive to the
configuration of the boundary of a catalyzing object.
The result in Fig. \ref{fig_app2} based on the more concrete set up
would be a supporting evidence for the universality of our main
proposal, that is, horizonless objects would catalyze vacuum decays
when its size is comparable with the size of a CDL bubble and
its compactness, $c \equiv \xi/2GM_{\text{tot}}$,
is of the order of unity, $c \sim {\mathcal O} (1)$.

%

\end{document}